\documentclass[preprint]{aastex}
\shorttitle{Super-Earth formation of extra solar planets}

\shortauthors{Ogihara et al.}

\title{ECCENTRICITY TRAP: TRAPPING OF RESONANTLY INTERACTING PLANETS NEAR THE DISK INNER EDGE}

\author{Masahiro Ogihara}
\affil{Tokyo Institute of Technology,
Ookayama, Meguro-ku, Tokyo 152-8551, Japan}
\email{ogihara@geo.titech.ac.jp}
\author{Martin J. Duncan}
\affil{Queen's University, Kingston, K7L 3N6, Ontario, Canada}
\email{duncan@astro.queensu.ca}
\and 
\author{Shigeru Ida}
\affil{Tokyo Institute of Technology,
Ookayama, Meguro-ku, Tokyo 152-8551, Japan}
\email{ida@geo.titech.ac.jp}

\begin{document}
\begin{abstract}
Using orbital integration and analytical arguments,
 we have found a new mechanism 
(an ``eccentricity trap'') to halt 
type I migration of planets 
near the inner edge of a protoplanetary disk.
Because asymmetric eccentricity damping due to disk-planet 
interaction on the innermost planet at the disk edge 
plays a crucial role in the trap, this mechanism
requires continuous eccentricity excitation
and hence works for a resonantly interacting convoy of planets.
This trap is so strong that the edge torque exerted on the
innermost planet can completely halt type I migrations of 
many outer planets through mutual resonant perturbations.
Consequently, the convoy stays outside the disk edge,
as a whole.
We have derived semi-analytical formula for the condition 
for the eccentricity trap and predict how many planets 
are likely to be trapped.
We found that several planets or more should be trapped
by this mechanism in protoplanetary disks that have cavities.
It can be responsible for the
formation of non-resonant, multiple, close-in super-Earth
systems extending beyond 0.1AU.  Such systems are being
revealed by radial velocity observations to be quite common around solar-type stars. 
\end{abstract}

\section{INTRODUCTION}
Recent radial velocity surveys indicate that a
significant fraction (40--60\%) of solar-type stars may harbor
close-in super-Earths, with masses and periods up to $\sim 20 M_\oplus$ 
and up to a few months, respectively (\citealt{mayor_etal09}; \citealt{bouchy_etal09}; 
\citealt{locurto_etal10}).
It is suggested that in many cases, these super-Earths 
are members of multiple-planet systems,
in which dynamical interactions may have influenced their formation
although their orbits would not be in mean-motion resonances 
in most systems.
Because mm radio observations suggest that the averaged disk mass 
around solar-type stars is comparable to the mass of 
the minimum-mass Solar nebula \citep{hayashi81}, the total mass of planetesimals
would not be enough for {\it in situ} accretion of the 
super-Earths in the innermost regions of disks.
They may have been accumulated by rocky planets that have
migrated from outer regions to the proximity of the host stars.
Because type I migration is halted at the inner disk edge, 
it may be reasonable that
super-Earths accrete near the edge.
If the inner disk edge is identified as accumulation
locations of hot jupiters, its radius may be $\sim 0.03$--0.05AU.
However, the observation suggests that averaged
orbital radius of the super-Earths is $\sim 0.1$AU,
well beyond the disk edge.
  
\citet{ogihara09} performed \textit{N}-body simulation to
study the accretion of planets from planetesimals near the disk edge.
Inward type I migration of planets is halted either by 
truncation of gas at the disk edge or 
by resonant perturbation from an inner planet. 
They found that in the case of 
relatively slow type I migration,
20--30 planets are captured by mutual mean-motion resonances
and these planets start orbit crossing 
and giant impacts after disk gas depletion,
resulting in the formation of several close-in super-Earths.
The super-Earths thus formed are kicked out of resonances
by strong scattering and collisions. 
This is in contrast to the fast migration case in which
only several planets survive during the presence of the gas
and they remain in stable resonant
orbits even after disk gas is removed (\citealt{terquem07};
\citealt{ogihara09}).

\citet{ogihara09} also found that in the slow migration case,
the innermost planet is pinned to the disk inner edge 
(set at $\sim 0.05$AU) and the convoy of the resonantly trapped 
planets extends well beyond 0.1AU.
Consequently, the super-Earths which form through the giant impacts
are distributed at $\sim 0.05$--0.3AU.
Thus, the resultant super-Earths are multiple, non-resonant
systems at $\sim 0.1$AU, which may be consistent with
observed data.

A question for the result of their \textit{N}-body simulation
is why the resonantly trapped convoy
of planets as a whole remain outside the disk edge.
Because individual planets (except the innermost one)
are losing angular momentum 
through type I migration and the angular momentum is redistributed
throughout the convoy with resonant interactions,
large amount of angular momentum must be supplied 
to prevent the planets from penetrating the disk edge.
Reverse torque of type I migration 
near the disk edge \citep{masset_etal06} can halt planets.
However, \citet{ogihara09} found the
trapping of a convoy outside the disk edge 
even in the case without the introduction of the reverse torque
and the effect of the reverse torque does not
change the efficiency of the trapping and
orbital configurations of trapped planets.

Here, in order to investigate the trapping mechanism,
we simulate in detail the orbital evolution of two planets 
(in some runs, more planets)
with various parameters for the disk edge,
the planet masses, and damping of eccentricities
and semimajor axes due to disk-planet interactions.
Together with analytical arguments, we find that
asymmetric damping of eccentricity of the innermost planet
near the disk edge is responsible for 
angular momentum input from the disk to the planet
with such a high rate that it compensates for the
angular momentum loss of all the trapped planets
due to inward type I migration.
We call this trapping mechanism an ``eccentricity trap.''

Although the eccentricity trap does not occur for
some parameter values of the disk and planets,
once the eccentricity trap occurs, it may be stronger
than the trap by the reverse type I torque. 
We derive semi-analytical formula for
the conditions necessary to produce an eccentricity trap, as a function
of planet masses, eccentricity and semimajor axis
damping timescales, and disk edge width.
We also predict how many planets can be trapped.
For likely parameter values for the innermost regions
in protoplanetary disks, the condition
for an eccentricity trap may be satisfied, 
while some systems including our solar
system might not have had a clear inner cavity in the disk
and therefore lost the inner planets.

In section~\ref{sec:ecc}, we present analytical argument 
to clearly show the intrinsic physics of the eccentricity trap.
In section~\ref{sec:num}, we show the results of \textit{N}-body simulations.
Comparing analytical arguments and the numerical results,
we derive quantitative conditions for the occurrence 
of the eccentricity trap.
In section~\ref{sec:con}, we summarize the results
and discuss their implications.

\section{ECCENTRICITY TRAP}
\label{sec:ecc}
\subsection{Basic Idea}
\label{sec:idea}
\begin{figure}
\epsscale{0.3}
\plotone{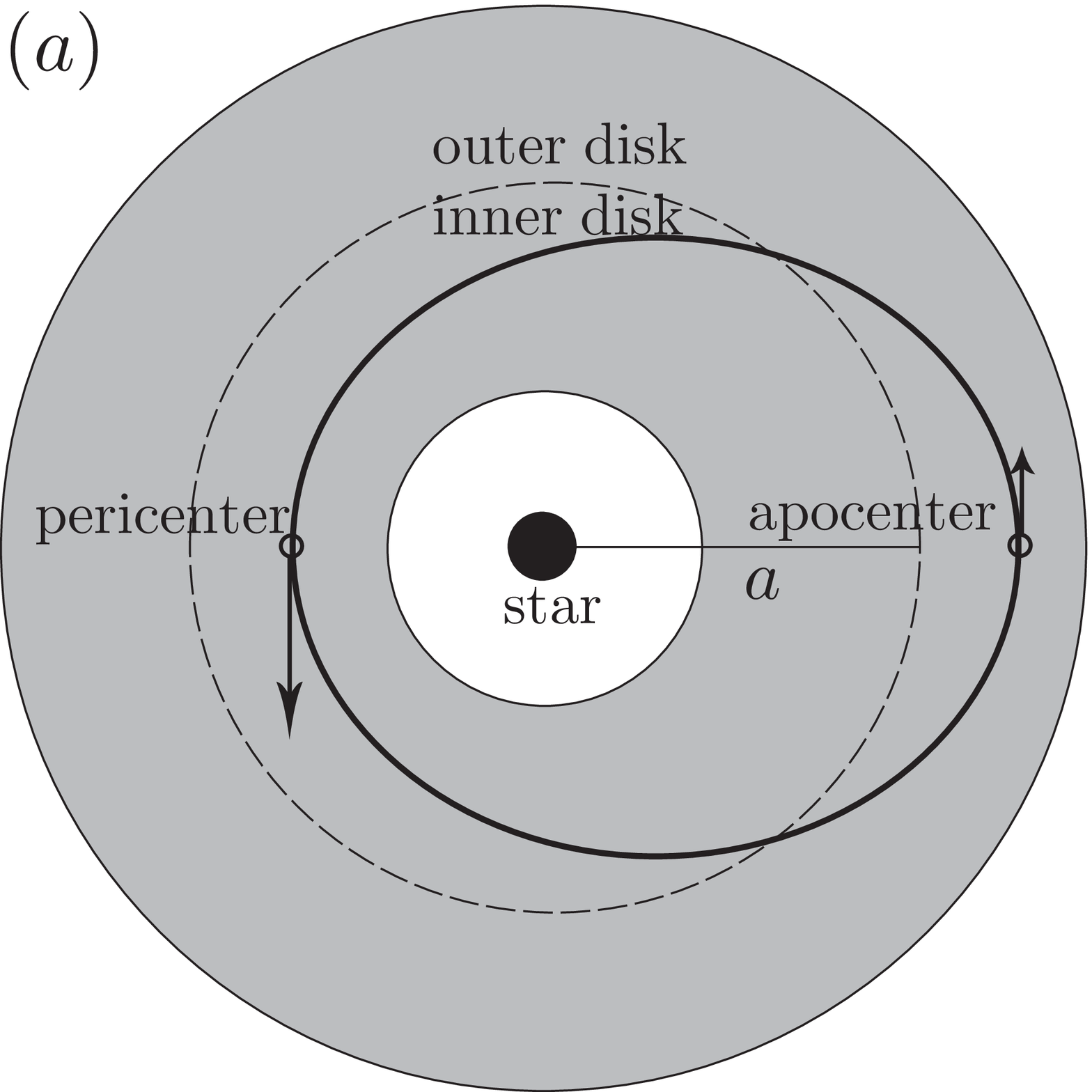}
\plotone{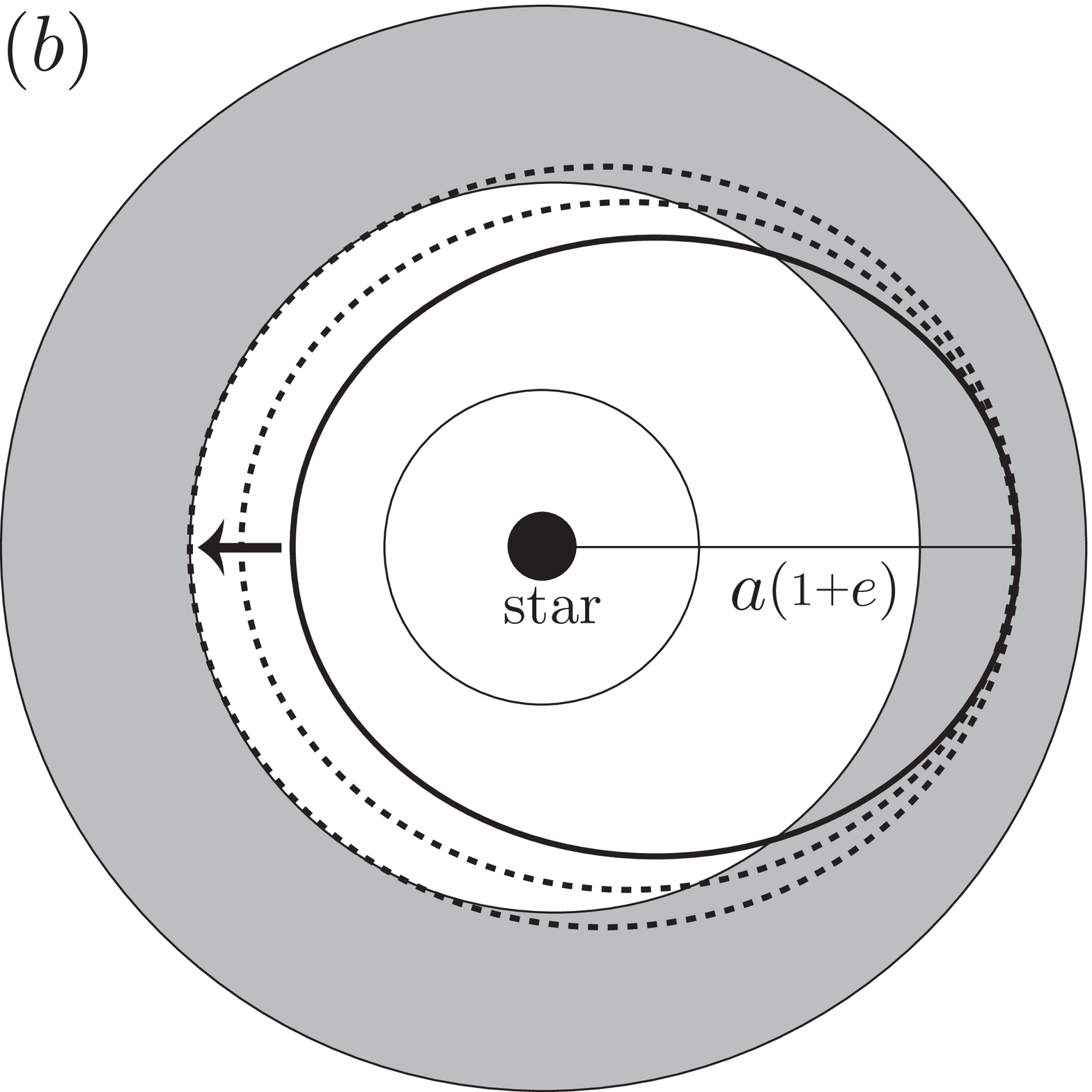}
\plotone{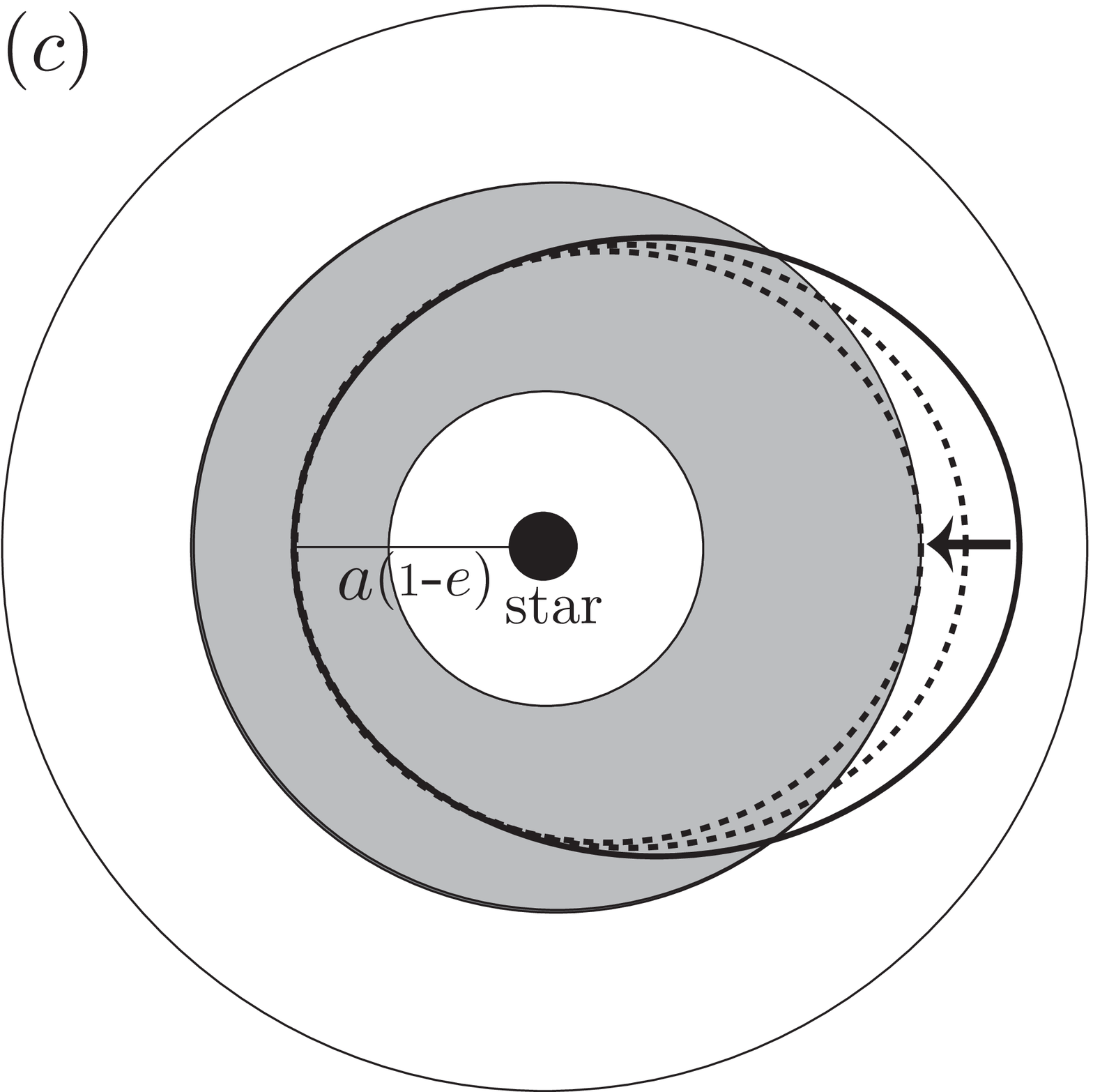}
\caption{Schematic illustration of expansion/shrink of $a$ due to $e$-damping.
(a) The solid line represents an eccentric orbit with semimajor axis $a$ 
around a star. The shaded toroidal region represents a gas disk. 
The guiding center, which divides the disk into two areas, 
is expressed by the dashed line.
At apocenter, the tangential velocity of the planet is slower than the
local gas velocity, while the orbital velocity is faster than the gas velocity 
at pericenter. 
(b) The planet suffers a tailwind when it is moving in the outer disk, 
which increases $a$ while decreases $e$
with the apocenter $a(1+e)$ kept almost constant.
(c) In the inner disk region, the planet suffers a headwind, leading to a
decrease of $a$, with the pericenter $a(1-e)$ kept almost constant.}
\label{fig:fig0}
\end{figure}

Consider a planet in an orbit
with eccentricity $e$ and semimajor axis $a$
in a gas disk (Fig.~\ref{fig:fig0}a).
We now consider changes in $a$ and $e$
caused by gravitational drag (dynamical friction)
from disk gas (\citealt{ostriker99}, \citealt{tanaka04};
for details, see section~\ref{sec:condition}).
The tangential velocity of the planet near the apocenter 
in an eccentric orbit is slower than local circular Keplerian
velocity when the planet moves through an outer disk region. 
Since the gas velocity is almost equal to circular 
Keplerian velocity, the planet suffers a tailwind. 
Then, $a$ increases while $e$ decreases, in such a way that $a(1+e)$
is almost kept constant (see schematic illustration in Fig.~\ref{fig:fig0}b).
On the other hand, 
the planet suffers a headwind when it moves through
an inner disk region.
Then, both $e$ and $a$ decrease such that $a(1-e)$
is almost kept constant (Fig.~\ref{fig:fig0}c).
Thus, $e$ is decreased by interactions with both
outer and inner disks, while changes in $a$ due to the
the outer and inner disks cancel out.

If the planet has an eccentric orbit that straddles
a relatively sharp disk edge inside which disk gas is depleted
to create a cavity, it suffers a drag force only 
when it moves in the outer disk region, so that
$e$ decreases and $a$ increases on orbit average.
Since the specific angular momentum of the planet is
\begin{equation}
L = \sqrt{GM_*a(1-e^2)},
\label{eq:L}
\end{equation}
where $M_*$ is the host star mass, 
the planet apparently gains orbital angular momentum
through gas drag. 

Note that type I migration torque decreases
$a$ even in the case of a circular orbit (e.g., \citealt{ward86}, \citealt{tanaka_etal02}).
In that case, the planet loses $L$ mostly by distant
perturbations from the outer disk and gains $L$
by those from the inner disk.
The loss and gain are almost balanced, but
the effects of curvature of the system and negative
pressure radial gradient slightly enhance the loss,
resulting in net inward migration (type I migration).
Because type I migration is caused by residual
of opposing torques while the $e$-damping does not have
such cancellation, the timescale of type I migration
($t_a$) is usually much longer than the 
$e$-damping timescale ($t_e$) (see section~\ref{sec:com}).

Since the $a$-increase by the edge effect
occurs on a timescale $\sim t_e$, 
the angular momentum gain rate due to the edge effect 
is much higher than the angular momentum
loss rate due to type I migration unless $e$ is extremely small (see Eq.~[\ref{eq:trap_cond0}]).
Therefore, if the planet migrates inward due to type I migration,
it should be ``trapped'' at the disk edge, as long as
orbital eccentricity is maintained to some level.
If another planet migrates to be trapped in a mean-motion
resonance with the inner planet trapped at the edge,
the resonant perturbations from the outer planet 
continuously pump up
eccentricity of the inner one and it enables
the eccentricity trap to be maintained.
Usually other planets also migrate inward and are captured at 
mutual mean-motion resonances one after another to form 
a convoy of planets.
Since the angular momentum gain due to the edge effect
is significant, it can compensate for the total loss of
several planets due to type I migration and trap
the convoy of planets as a whole outside the edge.

\subsection{Condition for the Trapping}
\label{sec:condition}
In this subsection, we derive an approximate 
condition for the eccentricity trap,
in order to understand the results of \textit{N}-body simulations
in section~\ref{sec:num}. 
Since in the \textit{N}-body simulations,
we integrate orbits by 
adding the instantaneous drag forces derived 
by \citet{tanaka04} to the equations of motion
(rather than secularly changing the orbital elements 
with orbit-averaged torques),
we derive the trap condition using their force formulas.

\citet{goldreich80}, \citet{ward88}, and \citet{artymowicz93}
derived the damping rate of $e$ due to planet-disk interactions.
They summed up resonances to obtain
an orbit-averaged $e$-damping rate, 
assuming uniformity of 
disk gas density on a scale of radial excursion of the planet.
In the situation we are considering, however, 
gas density that a planet passes through significantly 
changes during an orbital period, so that
we cannot apply their orbit-averaged results.
Derivation of the formulas for 
the torque exerted on a planet in an eccentric orbit
straddling a disk edge, based 
on full descriptions of Lindblad and corotation resonances, 
is left for future work.

On the other hand, \citet{tanaka04} derived
instantaneous drag forces caused by gravitational
potential perturbations due to density waves
at each point of epicycle motion of the planet.
They showed that the mean $e$-damping rate 
(Eq.~[\ref{eq:Tanaka_e}]) obtained by
averaging the instantaneous forces over one orbital period
agrees with that predicted by summing up resonant torques
in the case of locally uniform gas density.
\citet{ostriker99} calculated dynamical friction from
gas flow onto a particle. 
Since gas density is higher in downstream than in
upstream due to gravitational focusing, 
the particle is pulled back by the gravity of
the higher gas density, which is similar to Chandrasekahr's
dynamical friction in a stellar cluster \citep{binney87}.
\citet{tanaka04} also showed that spiral waves are
always stronger in backside of the epicycle motion
(Fig.~1 in their paper).
The Ostriker's force formula agrees with
Tanaka \& Ward's one except for a numerical factor.

Because the force formulas by \citet{tanaka04} are local
expressions, we adopt them both in \textit{N}-body simulations and
analytic arguments in this paper (see cautions below).
In the \textit{N}-body simulations, the forces are directly included
in equations of motion.  
The formulas for the specific damping forces are:
\begin{eqnarray} 
\textit{F}_{{\rm damp},r}  
   & = & \frac{1}{0.78 t_e} (2A^{c}_{r}[v_{\theta} - r \Omega] + A^{s}_{r} v_r) \label{eq:3-1} \\ 
\textit{F}_{{\rm damp}, \theta}  
   & = & \frac{1}{0.78 t_e} (2A ^{c}_{\theta}[v_{\theta} - r \Omega] + A ^{s}_{\theta} v_r) \label{eq:3-2}  \\ 
\textit{F}_{{\rm damp},z}  
   & = & \frac{1}{0.78 t_e} (A ^{c}_{z} v_z + A ^{s}_{z} z \Omega ) \label{eq:3-3} 
\end{eqnarray} 
where $v_r, v_{\theta}$, and $v_z$ are radial, tangential, and
vertical components of the planet's velocity, $r \Omega$ is
circular velocity of disk gas, which is almost equal 
to local Keplerian velocity, 
and the numerical factors are given by
\begin{eqnarray} 
A^{c}_{r} = 0.057 & & A^{s}_{r} = 0.176 \nonumber \\ 
A^{c}_{\theta} = -0.868  & & A^{s}_{\theta} = 0.325 \nonumber \\ 
A^{c}_{z} = -1.088  & & A^{s}_{z} = -0.871.\nonumber   
\end{eqnarray} 
In the case of locally uniform gas density,
integrating $\textit{F}_{{\rm damp},r}$ and 
$\textit{F}_{{\rm damp},\theta}$ over one orbital period, 
\citet{tanaka04} derived the averaged $e$-damping timescale ($t_e$),
\begin{equation}
t_e = \frac{e}{\dot{e}}  = 
\frac{1}{0.78}\left(\frac{M}{M_*}\right)^{-1} 
\left(\frac{\Sigma_g r^2}{M_*}\right)^{-1}
\left(\frac{c_s}{v_{\rm K}}\right)^{4} \Omega^{-1},
\label{eq:Tanaka_e}
\end{equation}
where $M$ and $M_*$ are masses of
the planet and the host star, 
and $\Sigma_g$ and $c_s$ are the gas surface density and
the sound velocity of the disk gas, respectively.

These Tanaka \& Ward's formulas assumed
subsonic flow.
We apply the formulas even for supersonic cases
in which radial excursion is wider than
the disk edge width that may be comparable to
disk scale height ($H$).  
\citet{ostriker99} showed that
the formula of the drag force must be
multiplied by a correction factor, 
$1/(1 + (v/c_s)^3)$, in the case of
supersonic flow.
The correction factor is consistent with a supersonic 
correction factor for an averaged $e$-damping rate
derived by \citet{papaloizou00},
if the relative velocity $v$ is identified by $e v_{\rm K}$.
We also carried out simulations with the 
correction factor (section~\ref{sec:supersonic}). 
Tanaka \& Ward's formulas also assumed
that gas density is locally uniform on
spatial scale of $\sim H$.
We apply the formulas also for the regions 
near the disk edge in which gas density 
may vary on a scale of $\sim H$.
The detailed analysis of its validity is left for future work,
because the purpose of the present paper is
to demonstrate a potential importance of the eccentricity
trap that we firstly found.

The edge effect described in section~\ref{sec:idea}
is quantitatively estimated 
with local (Hill) approximation, as follows.
In the Hill coordinates that are comoving with the planet's guiding
center at $a$, the tangential velocity of the planet at
$r = a + x = a (1 + e \cos (\Omega_0 t))$ is
\begin{equation}
v_y = - 2 e a \Omega_0 \cos (\Omega_0 t) = - 2 x \Omega_0,
\end{equation}
where $\Omega_0 = \sqrt{GM_*/a^3}$.
The Keplerian shear velocity 
(the local Keplerian circular velocity in the Hill coordinates)
of gas at $x$ is
\begin{equation}
v_{\rm shear} = - \frac{3}{2} x \Omega_0.
\end{equation}
The tangential relative velocity between the planet 
and local disk gas is
\begin{equation}
\Delta v = v_{\theta} - r \Omega = v_y - v_{\rm shear} 
= - \frac{1}{2} x \Omega_0.
\label{eq:del_v}
\end{equation}
Since the planet always moves slower than the gas
($\Delta v < 0$) at $x > 0$,
the drag force {\it accelerates} the planet's rotation there.
At apocenter, the above expression is also easily found
in global coordinates.
The tangential velocity of the planet at the apocenter 
in the inertial frame is
\begin{equation}
v_{\theta,{\rm apo}} = \frac{L}{a(1+e)} = \sqrt{\frac{GM_*(1-e)}{a(1+e)}},
\end{equation}
where $L$ is specific angular momentum of the planet
($L = \sqrt{GM_* a (1-e^2)}$),
while the local circular Keplerian velocity is
\begin{equation}
v_{\rm K,apo} = \sqrt{\frac{GM_*}{a(1+e)}}.
\end{equation}
Hence, for $e \ll 1$,
\begin{equation}
\Delta v = v_{\theta,{\rm apo}} - v_{\rm K,apo}
\simeq - \frac{1}{2} ea \Omega_0,
\end{equation}
which is identical to Eq.~(\ref{eq:del_v}) at the apocenter ($x = ea$). 

If the disk edge width is sharp enough
($ea \gg \Delta r$) and the edge is at $x \simeq 0$, 
the torque due to gravitational drag operates 
only at $x>0$, then the torque due to the $e$-damping
averaged over Keplerian time ($T_{\rm K} = 2 \pi /  \Omega_0$),
which we call ``edge torque,'' is
\begin{equation}
N_{\rm edge} \simeq \frac{1}{T_{\rm K}} 
\int_{-T_{\rm K}/4}^{T_{\rm K}/4} M r F_{{\rm damp},\theta} d t,
\label{eq:del_v0}
\end{equation}
where $M$ is the planet mass.
Since $v_\theta - r \Omega = (ae/2) \Omega_0 \cos(\Omega_0 t)$
and $v_r = ae \Omega_0 \sin(\Omega_0 t)$, 
Eq.~(\ref{eq:3-2}) becomes
\begin{equation}
F_{{\rm damp},\theta} = \frac{ae \Omega_0}{0.78 t_e}
\left(- A_{\theta}^c  \cos(\Omega_0 t)
+ A_{\theta}^s \sin(\Omega_0 t)\right).
\end{equation}
Then, Eq.~(\ref{eq:del_v0}) is reduced to
\begin{equation}
N_{\rm edge} 
\simeq - \frac{M}{T_{\rm K}} \int_{-T_{\rm K}/4}^{T_{\rm K}/4} \frac{A_{\theta}^c}{0.78} \frac{ea \Omega_0 \cos (\Omega_0 t)}{2 t_e} a d t
\simeq \frac{M}{2\pi} \frac{e a^2 \Omega_0}{t_e} 
\simeq \frac{e M L}{2\pi t_e} \; \; \left(= \frac{N_{e,{\rm apo}}}{\pi}\right),
\label{eq:N_edge_int}
\end{equation}
where $N_{e,{\rm apo}}$ is the torque
at the apocenter ($N_{e,{\rm apo}} \simeq e M a^2 \Omega_0/2t_{e}$).

If $\Delta r \gg ea$,
the planet suffers an opposing torque at $x < 0$ 
and the net torque must almost vanish on orbit average.
(In other words, with the integral range  
from $-T_{\rm K}/2$ to $T_{\rm K}/2$,
the above integral vanishes.)
Thus, in general cases, the edge torque can be 
expressed with a reduction factor $f$ $(0 < f < 1)$, which is a 
function of $ea/\Delta r$, as
\begin{equation}
N_{\rm edge} \simeq f \frac{e M L}{2\pi t_{e}}.
\label{eq:N_edge_max}
\end{equation}
If the edge is at $x > 0$, 
the range of integral in Eq.~(\ref{eq:N_edge_int})
is $-t_{\rm int}$ to $t_{\rm int}$ 
with $t_{\rm int} < T_{\rm K}/4$.
If the edge is at $x < 0$,
$t_{\rm int} > T_{\rm K}/4$.
In either case, the integrated value is smaller than
that in Eq.~(\ref{eq:N_edge_int}), which means that
Eq.~(\ref{eq:N_edge_max}) is the maximum value of
$N_{\rm edge}$.
 
In our \textit{N}-body simulation, 
even for planets near the edge, 
we apply the conventional inward migration
with a timescale given by \citep{tanaka_etal02}
\begin{equation}
t_a = - \frac{a}{\dot{a}} = \frac{1}{2.7+1.1q} \left(\frac{M}{M_*}\right)^{-1}
\left(\frac{\Sigma_g r^2}{M_*}\right)^{-1}
\left(\frac{c_s}{v_{\rm K}}\right)^{2} \Omega^{-1},
\label{eq:Tanaka_a}
\end{equation}
where $\Sigma_g \propto r^{-q}$.
\citet{masset_etal06} pointed out that near the
disk edge, due to the local positive
surface density gradient, the migration may be reversed 
to be outward.
The radiative effect can also make the migration outward
in optically thick disks
(e.g., \citealt{paardekooper_etal10}; \citealt{lyra_etal10}).
Furthermore, \citet{papaloizou00} suggested
that migration is slowed and reversed 
in supersonic cases (the migration timescale
has an additional factor of 
$(1 + (e v_{\rm K}/c_s)^5)/(1 - (e v_{\rm K}/c_s)^4)$).
Thus, our prescription for type I migration {\it underestimates} 
the efficiency of the trapping at the edge.
Nevertheless, our \textit{N}-body simulation shows that
the eccentricity trap does occur.
The effects of the local positive
surface density gradient at the disk edge and
radiation are left to future work.

In the equilibrium state of the trapping,
the edge torque is balanced with the sum of the type I 
migration torque on the planet at the edge 
and the resonant torque from an outer perturber ($- N_p$).
As we assume a conventional type I migration at the edge,
the type I migration torque on the innermost planet 
is given by $-M L/2t_a$.
Therefore, the condition for the planet
to be trapped at the edge is 
\begin{equation}
\frac{fe}{2 \pi}\frac{M L}{t_{e}} 
- \frac{M L}{2 t_a}
- N_p > 0.
\label{eq:dL_dt2}
\end{equation}

We first consider the simplest case of two equal-mass ($M$) planets.
In this case, the gravitational torque by the inner planet
on the outer planet, $-N_p$, balances 
the type I migration torque on the outer planet
given by $N_p \simeq M L/2 t_{a}$.
Then, Eq.~(\ref{eq:dL_dt2}) becomes
\begin{equation}
\frac{t_e}{t_a} < \frac{fe}{2\pi}.
\label{eq:trap_cond0}
\end{equation}
Here, we assume that difference in semimajor axes
of inner and outer planets is small enough and
represent them as a mean value, $a$,
in Eq.~(\ref{eq:trap_cond0}) for simplicity.
Thereby, although we consider trapping at the edge,
the trapping condition is reduced to
that parameterized by
the orbit-averaged damping timescales of $e$ and $a$
in uniform gas ($t_e$ and $t_a$).

\section{NUMERICAL SIMULATION}
\label{sec:num}
\subsection{Comparison between Theory and Simulation}
\label{sec:com}
The above trapping mechanism is quantitatively well demonstrated 
by simple two-planet systems in which the inner planet is 
located at a sharp disk edge. We have calculated the orbital 
evolution of two planets including 
effects of type I migration, gravitational drag, and the disk edge. 
The orbits of the planets are integrated by a 4th order Hermite scheme, 
adding instantaneous drag forces 
Eqs.~(\ref{eq:3-1}), (\ref{eq:3-2}), and (\ref{eq:3-3})
corresponding to $e$-damping and a tangential force,
\begin{equation}
\textit{F}_{{\rm mig}, \theta} = \frac{v_{\rm K}}{2 t_a},  
\label{eq:mig_f} 
\end{equation}
corresponding to $a$-damping, following \citet{ogihara09}. 

The parameters in the trapping condition (Eq.~[\ref{eq:trap_cond0}])
are $t_e/t_a$ and $\Delta r/r_{\rm edge}$,
where $r_{\rm edge}$ is the location of the disk edge.
If $\Delta r$ is comparable to the disk scale height ($H$),
\begin{equation}
\frac{\Delta r}{r_{\rm edge}} \simeq \frac{c_s}{v_{\rm K}}.
\end{equation}
(Note that the disk edge may be truncated by
stellar magnetic field and $\Delta r$ is not
necessarily $\sim H$.)
From Eqs.~(\ref{eq:Tanaka_e}) and (\ref{eq:Tanaka_a}),
\begin{equation}
\frac{t_e}{t_a} = \frac{2.7+1.1q}{0.78} \left(\frac{c_s}{v_{\rm K}}\right)^{2} \ll 1.
\end{equation}
At $r \sim 0.05$AU corresponding to the inner disk edge,
$T \simeq 1250$K in the optically thin limit \citep{hayashi81}.
Since $v_{\rm K} \simeq 130(r/0.05{\rm AU})^{-1/2}$km/s
and $c_s \simeq 2(T/1250{\rm K})^{1/2}$km/s,
$c_s/v_{\rm K} \simeq 0.016$ at $r \sim 0.05$AU.
In an optically thick disk, $T$ can be higher.
However, since silicate dust grains sublimate
and the disk becomes optically thin
at $T \ga 1400$K,
$T$ cannot be higher than that.
On the other hand, $t_a$ can be longer due to non-linear effects.
Hence, at $r \sim 0.05$AU,
$t_e/t_a \simeq 10^{-4}$--$10^{-3}$ and
$t_e \sim 100 T_{\rm K}$ for $M=M_{\oplus}$.
We perform calculations with wide ranges of $\Delta r/r_{\rm edge}$
and $t_e/t_a$ with fiducial values of $\Delta r/r_{\rm edge}=0.01$
and $t_e/t_a = 10^{-3}$.
In all runs, $t_e/T_{\rm K} = 100$ and $q=1.5$ are used, and
the gas surface density vanishes with a hyperbolic tangent
function of width $\Delta r$.

\begin{figure}
\epsscale{1.0}
\plotone{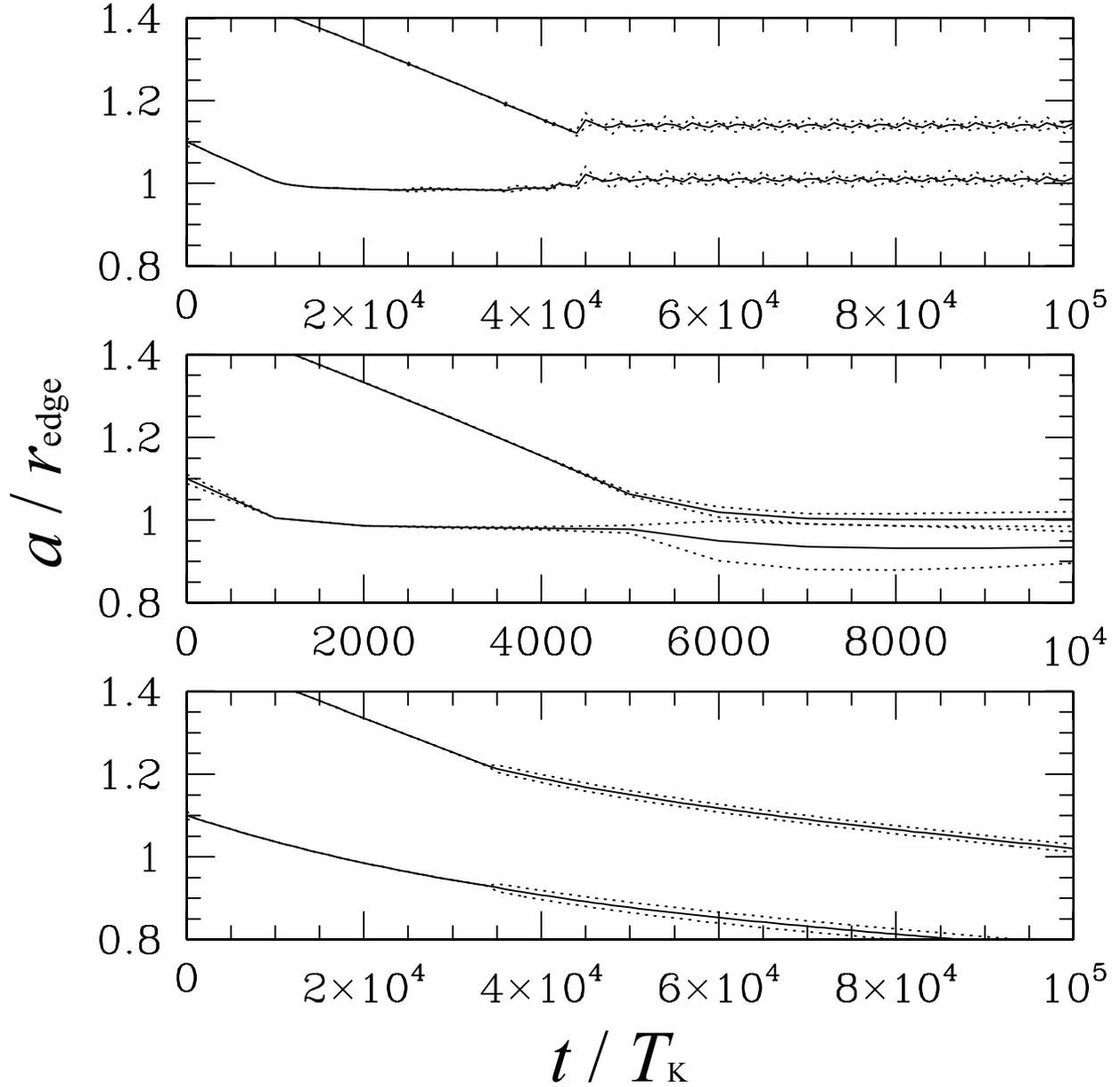}
\caption{Orbital evolution of the planets. Solid lines are semimajor axes, 
and dotted lines are apocenters and pericenters. 
\textit{Top:} The result of the fiducial case with $\Delta r/r_{\rm edge}=0.01$
and $t_e/t_a=10^{-3}.$
\textit{Middle:} The result of a fast migration case with $t_e/t_a=10^{-2}$.
\textit{Bottom:} The result of a smooth edge case with $\Delta r/r_{\rm edge}=0.2$.}
\label{fig:fig1}
\end{figure}

The results of the orbital integrations
are presented in Figure~\ref{fig:fig1}.
Two planets with Earth mass $M_\oplus$ are initially placed 
in almost circular orbits.
The top panel shows the orbital evolution of the fiducial case
with $\Delta r/r_{\rm edge} =0.01$ and $t_e/t_a = 10^{-3}$.
The solid lines represent the semimajor axis $a$, and the dotted 
lines represent the pericenters $a(1-e)$ and apocenters $a(1+e)$.
The inner planet migrates to the edge and stays there
because $\Sigma_g$ vanishes and $t_a \rightarrow \infty$ there.
After that, the outer planet migrates to be trapped in the 6:5 mean-motion 
resonance with the inner one.
Once the planets are captured in the resonance (at $t \sim 45000 T_{\rm K}$), 
the eccentricities of both planets are pumped up to $e\sim 0.01$
and their $a$'s are shifted slightly  outward, resulting in an
eccentricity trap.

The eccentricity trap depends on the parameter values.
The middle panel shows the result of fast migration
cases with $t_e/t_a=10^{-2}$ ($\Delta r/r_{\rm edge} =0.01$).
The planets are not trapped at the edge since the type I 
migration speed is too fast in this case, 
although the eccentricities become 
about 0.06 at the moment of capture into a resonance.
The bottom panel shows the result of
smooth edge case of $\Delta r/r_{\rm edge} =0.2$
($t_e/t_a = 10^{-3}$).
The planets also penetrate into the cavity because the
edge effect is weak.

We now derive a trapping condition by using the
simulation results. 
We calculated a torque due to
the gas drag at each timestep in the orbital integration 
and found that the orbit-averaged torque is
well fitted with 
\begin{equation}
f \simeq 
\left\{
\begin{array}{ll}
1 & (\mbox{for } e r_{\rm edge} > \Delta r ) \\
{\displaystyle 
\sqrt{\frac{e r_{\rm edge}}{\Delta r}}} & (\mbox{for } e r_{\rm edge} < \Delta r) .
\end{array}
\right.
\end{equation}
In the fiducial case with $\Delta r = 0.01r_{\rm edge}$, 
we found that $e \sim 0.01$, so that $f \simeq 1$.

\begin{figure}
\plotone{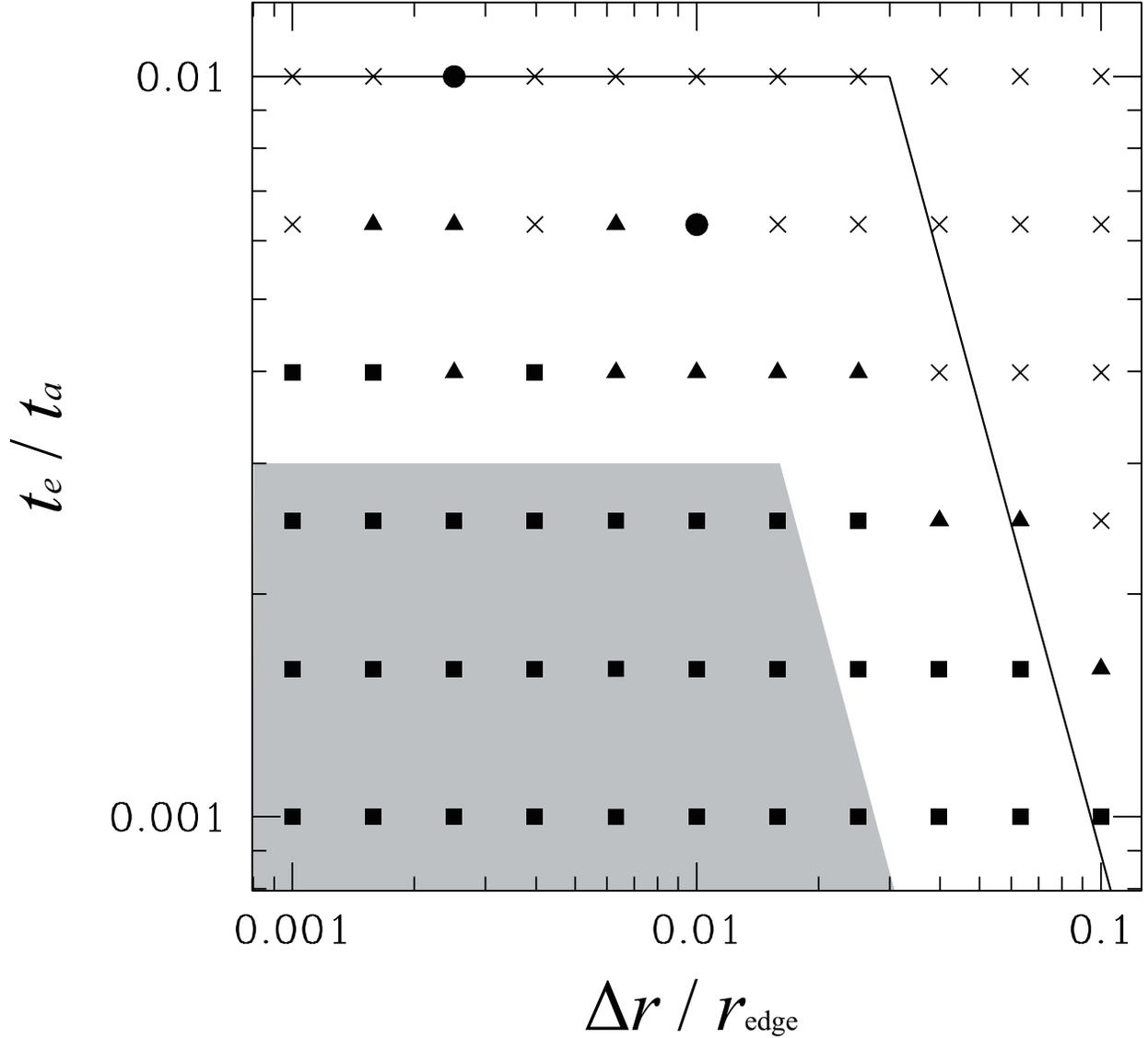}
\caption{
The results of trapping for various values of $\Delta r/r_{\rm edge}$
and $t_e/t_a$.
Crosses indicate that planets are not trapped, while other filled symbols 
indicate that planets are trapped. The filled squares, triangles, and circles 
represent the time-averaged eccentricity $e<0.02, 0.02<e<0.03$, and $e>0.03$, 
respectively.
The shaded region represents the theoretically predicted 
trapping regions (Eq.~[\ref{eq:condition}]). The solid line represents 
the upper limit (necessary condition) for the eccentricity trap.}
\label{fig:fig2}
\end{figure}

The results with various $t_{e}/t_{a}$ 
and $\Delta r/r_{\rm edge}$ for two Earth-mass planets
are summarized in Fig.~\ref{fig:fig2}.
Crosses represent the cases in which planets are not trapped.
Other symbols represent the trapped cases.
The filled squares, triangles, and circles
represent the time-averaged eccentricity of the inner planet
of $e<0.02$, $0.02<e<0.03$, and $e>0.03$, respectively.

\begin{figure}
\plotone{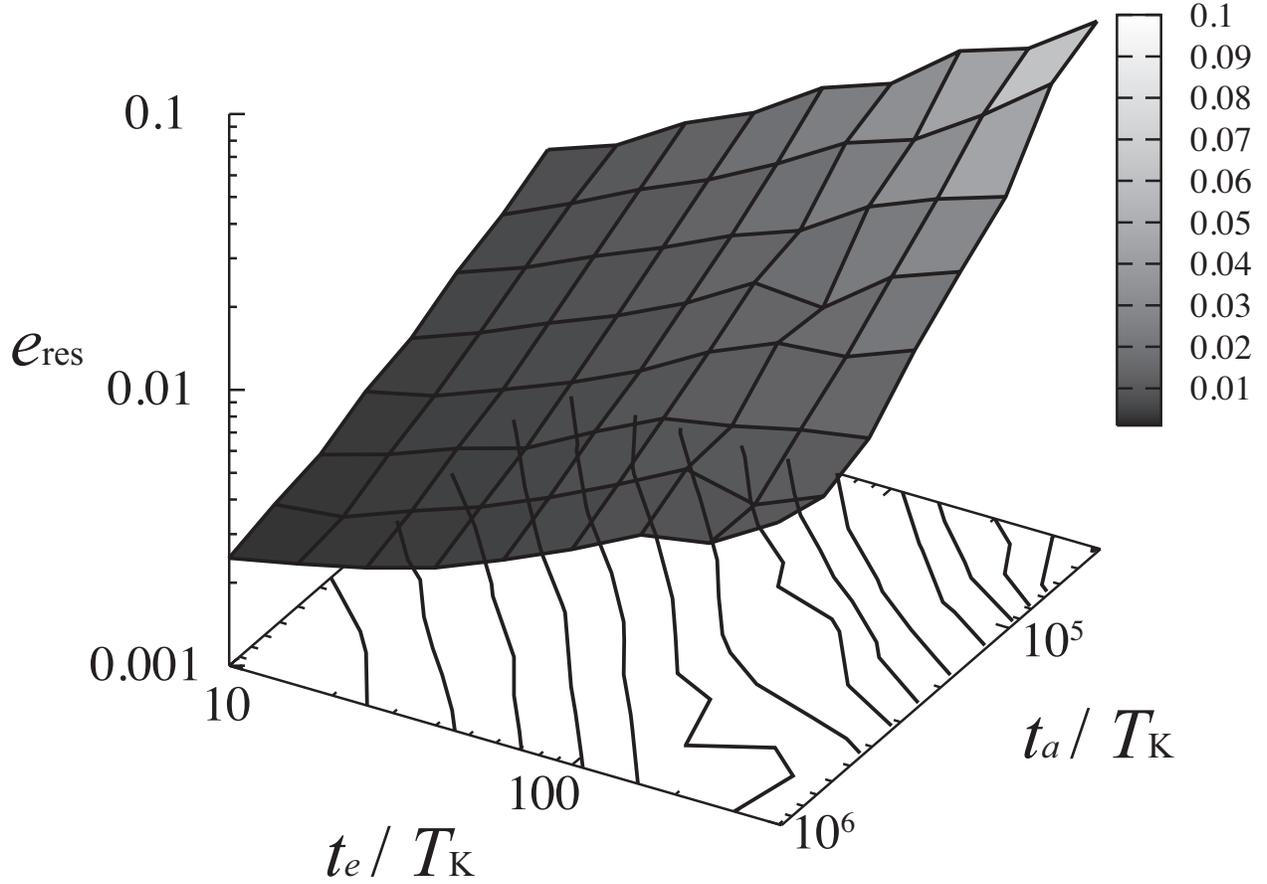}
\caption{
The relative eccentricity excited by
resonant trapping is plotted against $t_e$ and $t_a$ through numerical 
simulations. In these calculations,
$M_1 = M_2 =M_{\oplus}$ is assumed. 
The contours are drawn from $e_{\rm res} = 10^{-2.5}$  to $10^{-1.2}$
with intervals $\Delta \log_{10} e_{\rm res} =$ 0.1.
}
\label{fig:fig3}
\end{figure}

In order to compare the semi-analytical condition,
Eq.~(\ref{eq:trap_cond0}), with the orbital integration results,
we need a formula for resonantly
excited eccentricities.
We must also take into account the dependence on planetary masses.
Let $M_1$ and $M_2$ be the masses of the inner
and outer planets, respectively.
The eccentricity of the inner planet after resonant trapping
is expressed in the Hill approximation (e.g., \citealt{nakazawa88}) as
\begin{equation}
e \sim e_{\rm res} \frac{M_2}{M_1 + M_2} \equiv
       e_{\rm res} \frac{\nu_2}{2},
\label{eq:e_res1}
\end{equation}
where $e_{\rm res}$ is
the eccentricity of the relative motion 
(In the Hill approximation, relative motion between
two Keplerian motions is another Keplerian motion).
Since the magnitude of the relative eccentricity excited by
resonant interaction ($e_{\rm res}$) with damping 
is not theoretically clear, we use empirical values.
With $M_1=M_2=M_{\oplus}$, 
we have done the same calculations 
as Fig.~\ref{fig:fig2} 
for various values of $t_e$ [$=(10$--$10^{3})T_{\rm K}$]
and $t_a$ [$=(10^{4}$--$10^{6})T_{\rm K}$].
In Figure~\ref{fig:fig3}, $e_{\rm res}$ is plotted against $t_e$ and $t_a$. 
The relative eccentricity can be fitted by 
\begin{equation}
e_{\rm res} \simeq 0.02 
\left(\frac{t_e/t_a}{10^{-3}}\right)^{1/2}.
\label{eq:e_res}
\end{equation}
The trapping condition for a given
mean-motion resonance may be $t_{\rm lib} < t_a$, 
where $t_{\rm lib}$ is the libration timescale 
of the eccentricity by the secular perturbation,
\begin{equation}
t_{\rm lib} \sim \left(\frac{\mu}{M_*}\right)^{-1} 
\left(\frac{a_2}{a_1}\right)^3 T_{\rm K},
\end{equation}
where $\mu=M_1 M_2/(M_1+M_2)$ is the reduced mass,
$M_1$ and $M_2$ are masses of the inner and
outer planets, and $a_1$ and $a_2$ are their semimajor axes
(e.g., \citealt{murray99}). 
Since $t_{\rm lib}$ is larger for larger $a_2/a_1$, 
the planets tend to be trapped
by a more distant resonance for larger $t_a$.
As a result, $e_{\rm res}$ is smaller.
In section~\ref{sec:mass}, we will show 
through calculations with $M_1 = M_2 = (0.1$--$10)M_{\oplus}$
that the magnitude of $e_{\rm res}$ is 
almost independent of the planetary mass
in the trapped case as long as $t_e/t_a$ is fixed.
The fact that all of $t_{\rm lib}$, $t_e$ and $t_a$ are 
inversely proportional to planetary mass
suggests that the resonance at which planets are trapped
and hence the values of
$e_{\rm res}$ would not sensitively depend on planetary masses.
However, note that the formula 
(Eq.~[\ref{eq:e_res}]) is empirical and can strictly be
applied only for the range of parameters that we tested. 

Substituting $e = e_{\rm res}/2$ into Eq.~(\ref{eq:trap_cond0}),
the trapping condition is reduced to 
\begin{equation}
\frac{t_e}{t_a} \la
\left\{
\begin{array}{ll}
0.003 \nu_2^2
& \mbox{for } \Delta r/r_{\rm edge} < 0.017
\nu_2^{2}, \\
{\displaystyle
0.008 \nu_2^6
\left(\frac{\Delta r/r_{\rm edge}}{0.01}\right)^{-2}}
& \mbox{for } \Delta r/r_{\rm edge} > 0.017 
\nu_2^2,
\end{array}
\right.
\label{eq:condition}
\end{equation}
where $\nu_2 = 2M_2/(M_1 +M_2)$ and $\nu_2 = 1$ for
equal-mass planets. 
The shaded region in Fig.~\ref{fig:fig2} represents
the above semi-analytical trapping condition, which
is approximately consistent with the numerical results of orbital integrations.
Note that in \textit{N}-body simulations, 
the magnitude of the type I migration torque on the innermost
planet should be smaller than $ML/2t_a$ (which is adopted in 
Eq.~[\ref{eq:dL_dt2}]) because the innermost planet 
is located near the edge where gas density is smaller. Accordingly, 
the trapping region would be larger than the shaded region. 
On the other hand, neglecting 
the type I migration torque on the innermost planet in Eq.(\ref{eq:dL_dt2}), 
we obtain the necessary condition for trapping (solid line in 
Fig.~{\ref{fig:fig2}}). 
The transition from the trapping and non-trapping should occur 
in the region between the outer boundary of the shaded region 
and the solid line, which is consistent with the numerical results.
Thus, the condition Eq.~(\ref{eq:trap_cond0}) is a slightly conservative 
condition.
Since the semi-analytical condition 
includes planetary mass dependence through $\nu_2$, 
we next examine it
by comparing orbital integrations and the semi-analytical argument.

\subsection{Dependence on Planetary Mass}
\label{sec:mass}
The dependence of trapping condition on the planetary
mass is investigated in this subsection.
Let $M_1$ and $M_2$ be masses of the inner and outer planets.
As in the discussion on Eq.~(\ref{eq:dL_dt2}),
we represent their semimajor axes ($a_1$ and $a_2$) 
as a mean value, $a$ $(\sim a_1, a_2)$, for simplicity.
In this case,
Eq.~(\ref{eq:dL_dt2}) is
\begin{equation}
\frac{fe_1}{2 \pi}\frac{M_1 L}{t_{e}(M_1)} 
- \frac{M_1 L}{2 t_{a}(M_1)} - N_p > 0, 
\label{eq:dL_dt3}
\end{equation}
where
\begin{equation}
N_p = -\frac{M_2 L(a_2)}{2 t_{a}(M_2,a_2)} 
= -\frac{M_1 L(a_1)}{2 t_{a}(M_1,a_1)} \left(\frac{a_2}{a_1} \right)^{-q+1/2} 
\left(\frac{M_2}{M_1}\right)^2 
\simeq -\frac{M_1 L}{2 t_{a}(M_1)} \left(\frac{M_2}{M_1}\right)^2. 
\label{eq:grav_torque}
\end{equation}
With Eq.~(\ref{eq:e_res}), the trapping condition is given by
\begin{equation}
\frac{(M_1^2 + M_2^2) (M_1 + M_2)}{M_1^2 M_2} < C,
\label{eq:condition5}
\end{equation}
where
\begin{equation}
C = \frac{e_{\rm res}f}{\pi} \frac{t_a}{t_e}
= \frac{20f}{\pi}\left(\frac{t_e/t_a}{10^{-3}}\right)^{-1/2}.
\label{eq:condition5.5}
\end{equation}
For a fiducial case with $t_e/t_a = 10^{-3}$ and
$\Delta r/r_{\rm edge} = 0.01$,
$e_{\rm res} \simeq 0.02$ and $f \simeq 1$, 
so that $C \simeq 6$.

\begin{figure}
\plotone{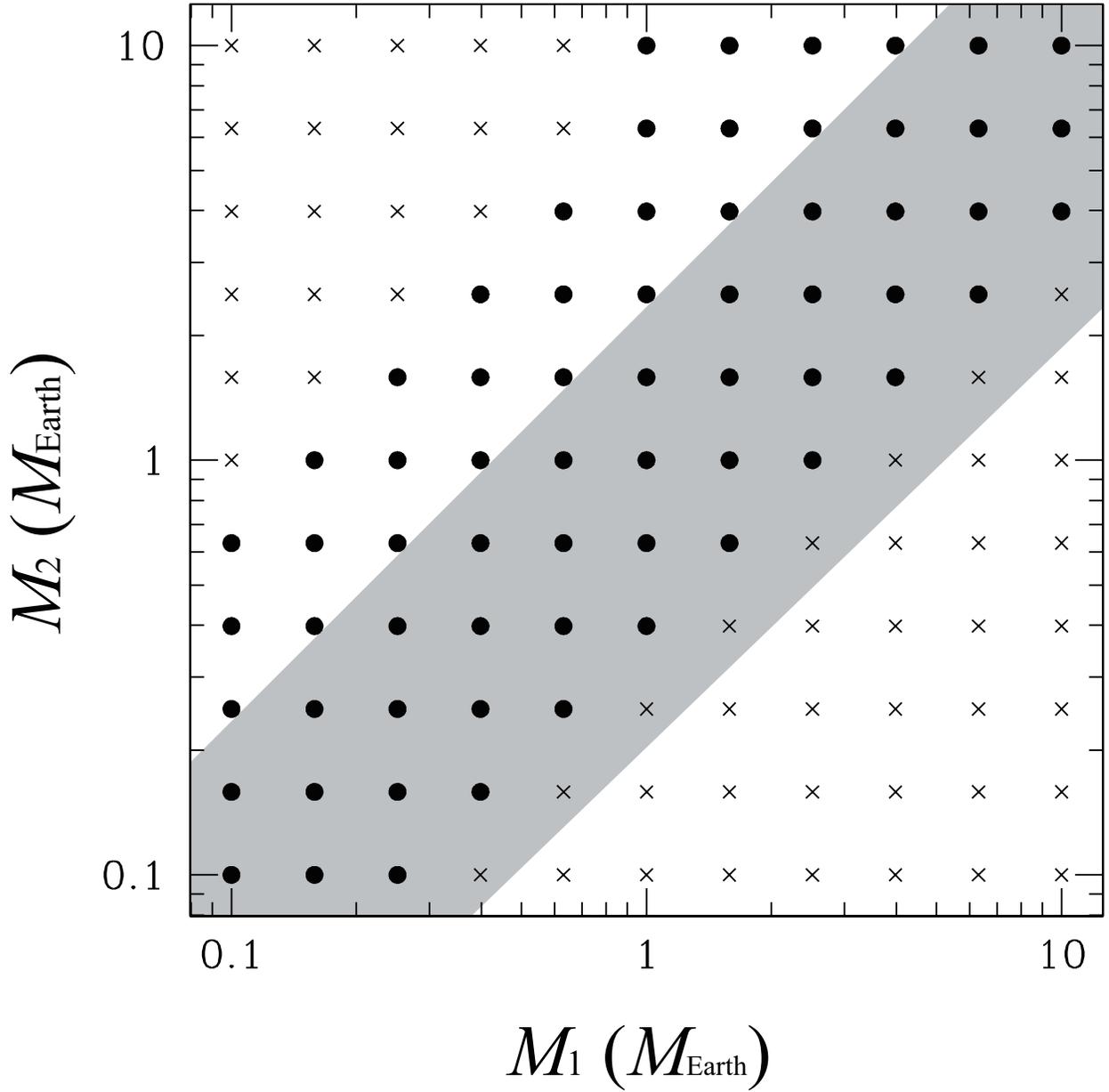}
\caption{The results of trapping for
various masses of inner and outer planets ($M_1$, $M_2$). 
Filled circles represent the cases where planets are trapped, while crosses 
represent non-trapped cases.
The shaded region represents the theoretically predicted
trapping regions given by Eq.~(\ref{eq:condition5}).
}
\label{fig:fig4}
\end{figure}

To confirm the above condition, 
we carried out additional numerical simulations with various planetary
masses, (0.1--$10)M_{\oplus}$.
We used the fiducial parameters, $t_e/t_a = 10^{-3}$ and
$\Delta r/r_{\rm edge} = 0.01$.
Since the $e$-damping time is set such that 
$t_e = 100(M/M_{\oplus})^{-1}T_{\rm K}$,
in order to keep $t_a/t_{\rm lib}$ constant,
we found that the resonance at which planets are trapped
is similar 
and $e_{\rm res} \sim 0.02$
in most of the trapped cases 
even if $M_1$ and $M_2$ are changed from 
$0.1M_{\oplus}$ to $10M_{\oplus}$.
Note that $e_{\rm res}$ becomes larger than 0.02
when $M_2 \sim CM_1 > M_1$.
Figure~\ref{fig:fig4} summarizes the results on the $M_1-M_2$ plane.
The trapped and non-trapped cases are indicated by
filled circles and crosses, respectively.
The shaded region represents the trapping condition (Eq.~[\ref{eq:condition5}]).
The results are consistent with the analytical condition
except that the trapping is more extended to high $M_2/M_1$ regions,
because \textit{N}-body simulations show higher $e_{\rm res}$ than
that assumed in the analytical estimate in the high $M_2/M_1$ regions.

The results might seem rather counter-intuitive.
That is, the trap does not occur when the outer planet is 
very small, while the trap does occur even if the outer planet is
somewhat more massive than the inner one. 
If the outer planet is too small,
the excited magnitude of $e_1$ is too small for the
edge torque to be effective.
On the other hand, even if the outer planet is more massive,
the trapping can occur, because $e_1$ is excited highly enough.
For a further massive outer planet, 
the edge torque is no more balanced with angular momentum loss
due to type I migration of the massive outer planet. 
Thus, the eccentricity trap works for
planets with comparable masses, which is often the case
during planet formation, because 
planets actually start
migration when migration timescale ($\propto M^{-1}$)
becomes shorter than growth timescale ($\propto M^{1/3}$)
and hence the migrating planets should have
similar masses.

\subsection{Number of Trapped Bodies}

We also investigated systems with more than two planets.
The trapping condition becomes more severe than in two planet
systems, since the edge torque on the innermost planet 
must balance the type I migration torques of all the planets.
The number of planets that can be trapped by the edge effect
increases with increase in the type I migration timescale. 
In fact, \citet{ogihara09} found that 20--30 planetary embryos 
are lined up by the edge 
effect in the case of slow migration. 

We estimate the maximum number of trapped planets.
For simplicity, we assume equal mass bodies.
When $n$ planets are trapped near the disk edge,
Eq.~(\ref{eq:grav_torque}) is replaced by
\begin{equation}
N_p \sim - \frac{n-1}{2 t_{a}}ML,
\end{equation}
where we also neglected the $a$-dependence, although
it may not be able to neglected for sufficiently large $n$.
Then, Eq.~(\ref{eq:dL_dt2}) becomes 
\begin{equation}
\frac{f e_{\rm res}}{4 \pi}\frac{1}{t_{e}} 
- \frac{1}{2 t_{a}}
- \frac{n-1}{2 t_{a}} > 0.
\label{eq:dL_dt4}
\end{equation}
Therefore,
\begin{equation}
n < \frac{f e_{\rm res}}{2 \pi} \frac{t_{a}}{t_{e}}.
\label{eq:condition4_0}
\end{equation}
Through \textit{N}-body simulation, 
we also derived the dependence of $n$ on $e_{\rm res}$ 
(Eq.~[\ref{eq:e_res}]) as
\begin{equation}
e_{\rm res} \simeq 0.02 
\left(\frac{t_e/t_a}{10^{-3}}\right)^{1/2}
\left(\frac{n}{2}\right)^{1/2}.
\end{equation}
Then, Equation~(\ref{eq:condition4_0}) becomes
\begin{equation}
n < \frac{50 f^2}{\pi^2} \left(\frac{t_e/t_a}{10^{-3}}\right)^{-1}.
\label{eq:condition4}
\end{equation}

\begin{figure}
\plotone{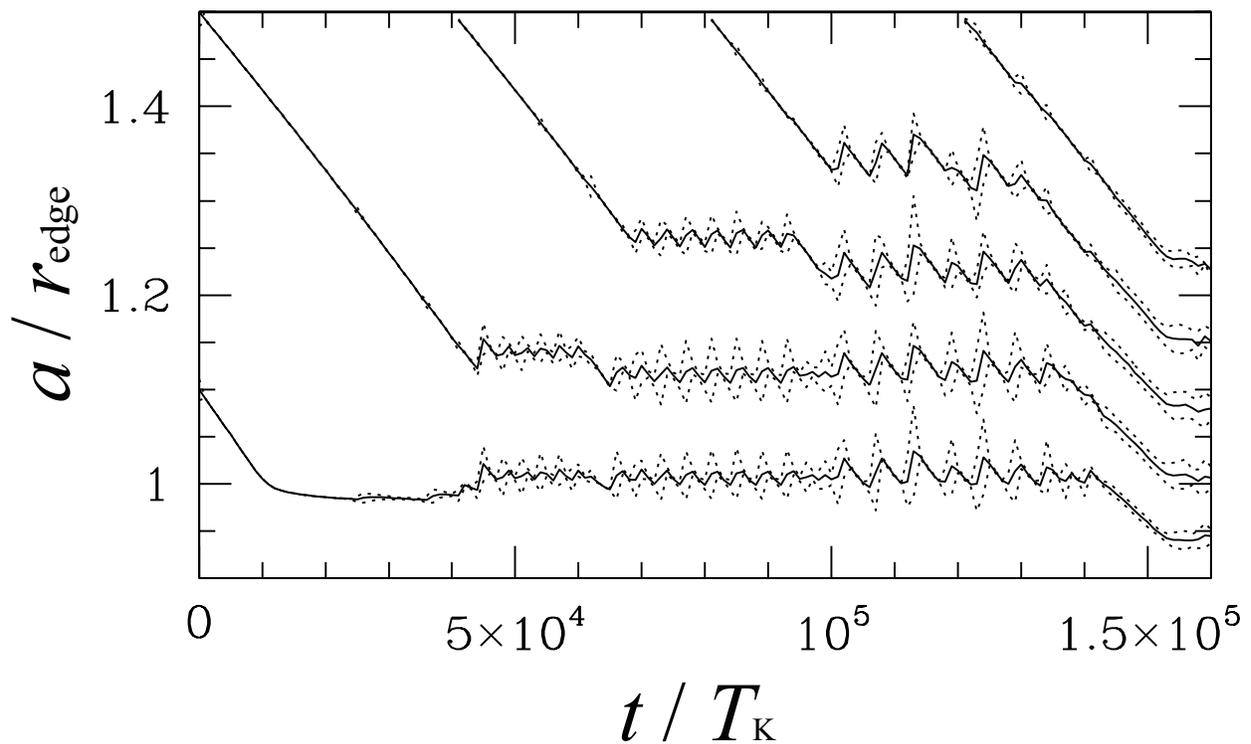}
\caption{Orbital evolution of multiple ($n > 2$) 
Earth-mass planets in the fiducial case 
($\Delta r/r_{\rm edge}=0.01$ and $t_e/t_a=10^{-3}$). Solid lines are 
semimajor axes, and dotted lines are pericenters and apocenters. 
}
\label{fig:fig5}
\end{figure}

To confirm Eq.~(\ref{eq:condition4}),
we performed a calculation with $n \ge 3$ in
the fiducial case with $\Delta r/r_{\rm edge} =0.01$ and 
$t_e/t_a=10^{-3}$.
In this case, Eq.~(\ref{eq:condition4}) is reduced to $n < 5$.
Individual planet masses are one Earth mass.
The orbital evolution is shown in Fig.~\ref{fig:fig5}.
After the first planet is trapped at the edge, 
subsequent migrating planets are trapped one after another.
After $10^5 T_{\rm K}$, four planets are
trapped by the edge torque.
But, when the fifth planet is trapped, the 
edge torque no longer halts the planets outside the
edge and the
they migrate inward as a whole.
The innermost planet is pushed into the cavity.
After $1.5 \times 10^5 T_{\rm K}$, four planets are
outside the edge and the edge torque exerted 
on the second planet supports the outer two planets.
The result of \textit{N}-body simulation
is in a good agreement with the analytical estimate, $n < 5$.

With the value of $t_e/t_a \sim 10^{-5}$ that \citet{ogihara09} used
in their slow migration case,
Eq.~(\ref{eq:condition4}) shows that $n < 500$.
This is consistent with their result in which
they found that all the (20--30) migrating planetary embryos
are trapped and remain outside the disk edge.

\subsection{Supersonic Correction}
\label{sec:supersonic} 
The Tanaka \& Ward's formulas assume subsonic flow. 
However, for the eccentricity trap to be efficient,
eccentricity $e$ must be pumped up by resonance to be
$e a \ga H$, that is, $e v_{\rm K} \ga c_s$.
In this case, the relative motion between gas and planets
can be supersonic.
According to \citet{ostriker99}, we also
carried out runs with 
the $e$-damping forces (Eqs.~[\ref{eq:3-1}]--[\ref{eq:3-3}])
multiplied by a supersonic correction factor, 
$1/(1+(ev_{\rm K}/c_s)^3)$.
Then, the effective damping timescale $t_{e,\rm{eff}}$ is defined as
\begin{eqnarray}
t_{e,\rm{eff}} \equiv t_e \times \left[1+\left( \frac{e v_{\rm K}}{c_s} \right)^3 \right],
\end{eqnarray}
where $t_e$ is the Tanaka \& Ward's damping timescale 
(Eq.~[\ref{eq:Tanaka_e}]).
This supersonic correction is consistent with that 
obtained by \citet{papaloizou00}.

Because $t_{e,\rm{eff}}$ is larger than $t_e$ for supersonic cases,
this correction is negative for the eccentricity trap.
However, \citet{papaloizou00} showed that
supersonic correction is also applied for the $a$-damping.
The correction factor for the $a$-damping timescale is
$(1 + (e v_{\rm K}/c_s)^5)/(1 - (e v_{\rm K}/c_s)^4)$,
which is a stronger function of $e v_{\rm K}/c_s$ than
that for the $e$-damping timescale and even changes sign to 
make the migration outward.
With the corrections for both $a$ and $e$ damping,
the eccentricity trap is rather strengthen in supersonic cases.
Here, we include the supersonic correction only for the $e$-damping
in \textit{N}-body simulations,
which significantly {\it underestimates} the efficiency of 
the eccentricity damping.
Nevertheless, the eccentricity trap still occurs for
likely values of $t_e/t_a$, although the region
of the eccentricity trap on the $\Delta r/r_{\rm edge}-t_e/t_a$ 
plane is more restricted. 

\begin{figure}
\plottwo{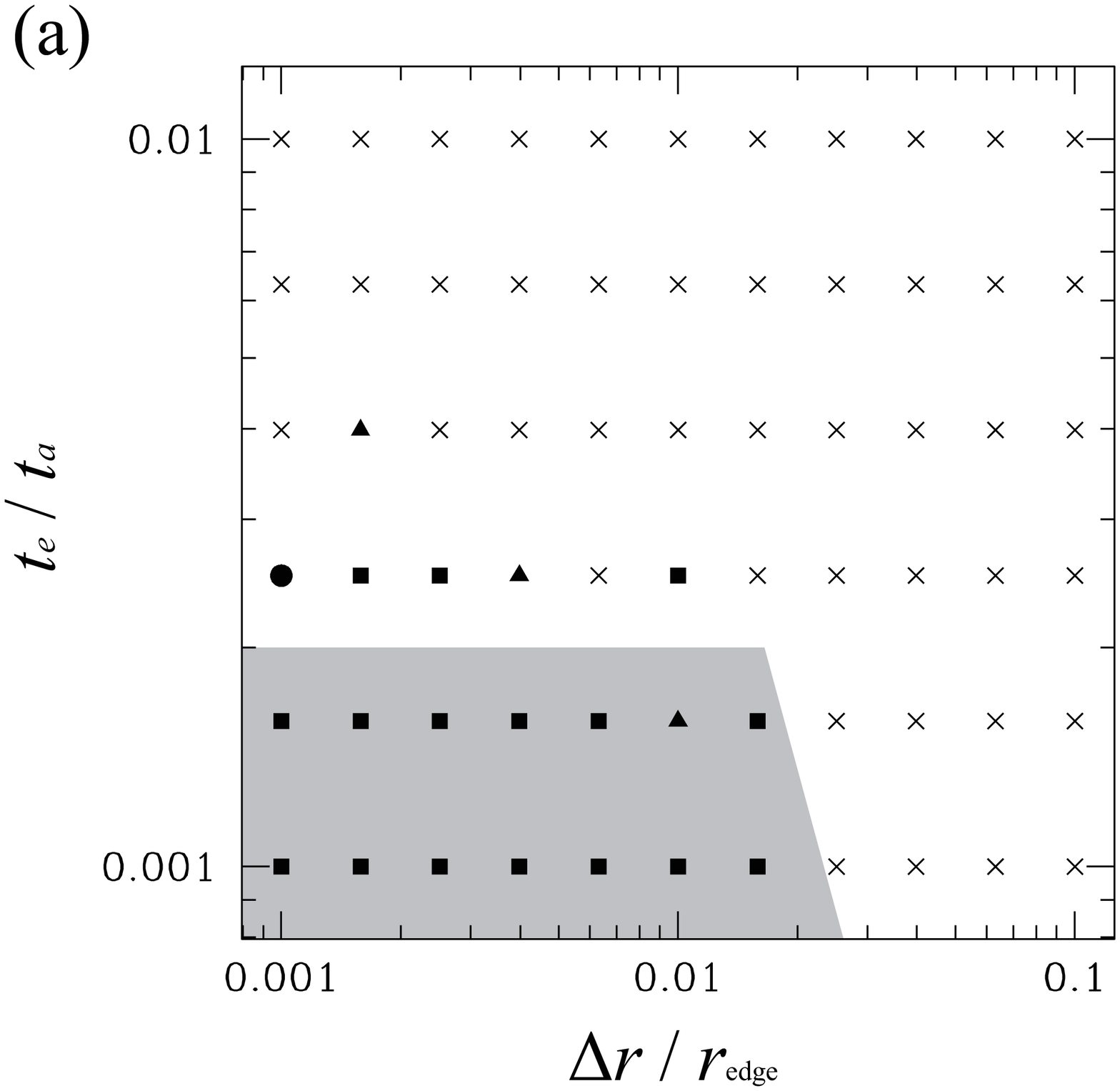}{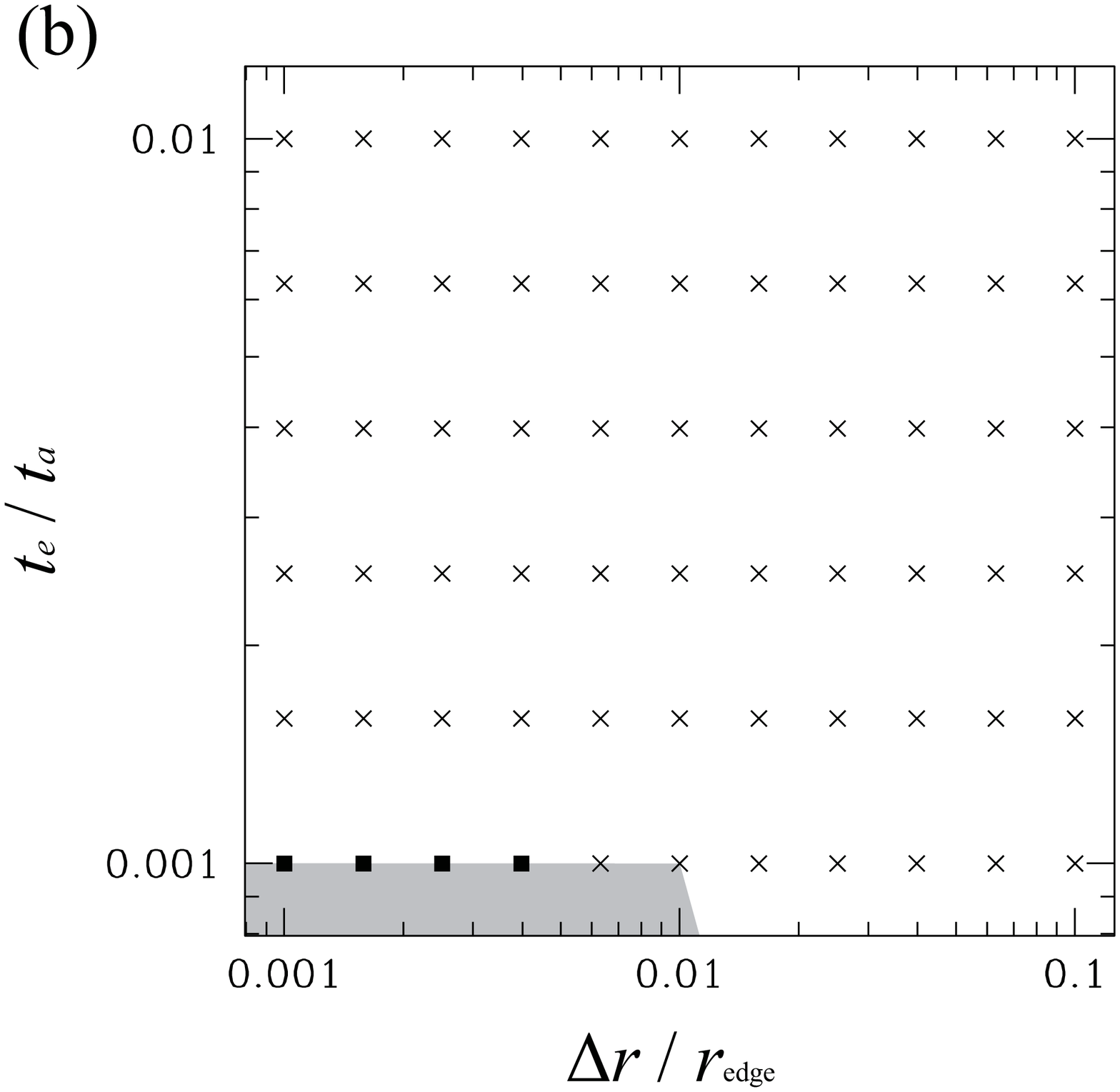}
\caption{Same as Fig.~\ref{fig:fig2}, but with supersonic correction
factor. (a) Case with $c_s/v_{\rm K} \simeq 0.03$. 
(b) Case with $c_s/v_{\rm K} \simeq 0.02$. The shaded region is 
the analytically predicted trapping regions. Note that $t_e$ is not 
the effective damping timescale.}
\label{fig:fig7}
\end{figure}

Figure~\ref{fig:fig7} summarizes 
the results with the supersonic 
correction on the $\Delta r/r_{\rm edge}-t_e/t_a$ plane.
(Note that $t_e$ which is used in the vertical axis is not the effective 
damping timescale but that for subsonic cases described in 
Eq.~[\ref{eq:Tanaka_e}].)
Figure~{\ref{fig:fig7}a} and {\ref{fig:fig7}b} are the cases with 
$c_s/v_{\rm K} \simeq$ 0.03 and 0.02, respectively.
While the eccentricity trap regions are more restricted, in particular,
in Figure~{\ref{fig:fig7}b}, they still exist. 
The shaded regions in Fig.~\ref{fig:fig7} represent
the (conservative) analytical condition (Eq.~[\ref{eq:trap_cond0}])
with the effective damping timescale, which are consistent with
the numerical results.

\section{CONCLUSION}
\label{sec:con}
We propose a mechanism that we call an ``eccentricity trap'' 
of migrating planets
and investigated the details of 
the trapping condition both analytically and numerically.
This halting mechanism is so strong that many resonantly-interacting
planets are
trapped by the edge torque that is exerted only on an
innermost planet at the inner disk edge, 
as long as the edge is sharp enough 
($\Delta r < r_{\rm edge}$; $\Delta r$ is the edge width
and $r_{\rm edge}$ is the edge radius.)
and the eccentricity damping timescale 
due to planet-disk interaction ($t_e$)
is short enough compared with 
semimajor axis damping timescale ($t_a$).
The explicit trapping condition is given by
\begin{equation}
\frac{t_e}{t_a} < \frac{fe}{2\pi},
\label{eq:cond_last}
\end{equation}
where
\begin{equation}
f \simeq 
\left\{
\begin{array}{ll}
1 & (\mbox{for } e r_{\rm edge} > \Delta r ) \\
{\displaystyle 
\sqrt{\frac{e r_{\rm edge}}{\Delta r}}} & (\mbox{for } e r_{\rm edge} < \Delta r) .
\end{array}
\right.
\end{equation}
The evaluation for the eccentricity of the innermost body, $e$,
is given by Eqs.~(\ref{eq:e_res1}) and (\ref{eq:e_res}).

As discussed in section~\ref{sec:com}, near the inner disk edge ($ \la 0.1$AU),
the sound velocity scaled by Kepler velocity is so small that
disk scale height may be sharp ($\Delta r/r_{\rm edge} \sim 10^{-2}$)
and $t_e/t_a$ is as small as $10^{-4}$--$10^{-3}$
even if we assume the formulas for the $a$ and $e$ damping
by \citet{tanaka_etal02} and \citet{tanaka04}.
Therefore, the eccentricity trap condition is safely satisfied
in protoplanetary disks, if they have a cavity.
When the motion of the planet is supersonic ($e v_{\rm K}>c_s$),
damping formulas are modulated.
Then, the eccentricity trap is strengthen.

We also derived a formula for the maximum number of 
trapped planets outside the edge.
It explains the result of \citet{ogihara09}, where 
20--30 planets are lined up outside the disk edge in the case 
of slow migration.
This is an essential point of \citet{ogihara09}'s model
for formation non-resonant, multiple, close-in super-Earths
at $\sim 0.1$AU.
The innermost planet is pinned to the disk inner edge 
(at $\sim 0.05$AU) and the convoy of the resonantly trapped 
planets extends well beyond 0.1AU.
After gas dispersal, they undergo orbital crossing and are 
kicked out of the resonances,
leading to many giant impacts to increase the planetary
mass by more than a factor of 10.
As a result, a few super-Earths are accreted
at $\sim 0.05$--0.3AU.
Since their masses exceed several Earth masses after
disk dissipation, they avoid the gas accretion necessary to become
gas giant planets, as we show in a sequential planet
formation model in a separate paper \citep{ida_lin10}. 
The resultant non-resonant, multiple super-Earths
systems at $\sim 0.1$AU are consistent with
observed data.
The discovered close-in super-Earth systems (GJ~581 and HD~40307) 
do not have any commensurate relationships, 
and orbits of outer planets extend beyond 0.1~AU,
which can be formed reasonably by the above model
if clear cavities existed in their progenitor disks.

This trapping mechanism also provides deep
insights into satellite formation around 
giant planets. 
Based on the slow inflow disk model (e.g., \citealt{canup02,canup06}),
$c_s/v_{\rm K} \sim 0.06$ and we found through
a preliminary \textit{N}-body simulation that $e \sim 0.1$
for resonantly trapped satellites.
Equation~(\ref{eq:condition4_0}) suggests that
a few Galilean satellites are trapped outside the disk edge. 
If more satellites are trapped to violate the trapping condition, 
the innermost one
is pushed into the inner cavity.
\footnote{
In the case of proto-satellite disks, the 
satellite mass cannot be neglected compared with
the disk mass.
When the total mass of trapped satellites 
exceeds the disk mass, the trapping may become no more 
effective as well.
This limit for trapping is similar to
the above argument of the maximum $n \sim$ a few 
\citep{sasaki_etal10}.
}
Since the disk edge would coincide with corotation radius
with the host object's spin, 
the innermost satellite in the cavity loses its orbital angular momentum
by tidal torque from Jupiter to collide with it.
As a result, resonant trapping of a few satellites
may be always maintained.
Because the number of the trapped satellites are relatively small,
the satellites may remain stable after gas dissipation.
An outermost satellite, Callisto, 
which is not in a resonance, may accrete in the outer
residual disk.
More detailed study of satellite formation 
is discussed in \citet{sasaki_etal10} and a forthcoming paper
 (Ogihara et al. in preparation).

The validity of applying the $e$-damping 
formulas by \citet{tanaka04} for nonuniform region near the disk edge
should be tested by hydrodynamical simulations.
At the same time, reverse torque for type I migration \citep{masset_etal06}, 
which is not considered here, should also be considered as well.
Although we simply assumed the disk edge width is
similar to the disk scale height, disk edge structure may be created
by stellar magnetic field.
Thus MHD simulations are also needed to determine 
the disk edge structure.
Furthermore, the conditions necessary to maintain a cavity should also
be investigated by MHD simulation.

\subsection*{ACKNOWLEDGMENT}
We thank fruitful and stimulating discussions with
Steve Lubow, Bill Ward, Doug Lin, John Papaloizou,
Aur\'{e}lien Crida, Clement Baruteau, and Sijme-Jan Paardekooper 
during our participation in ``Dynamics of Discs and Planets'' workshop
at the Newton Institute at Cambridge University, 
where some of this work was done.
We also thank the anonymous referee for helpful comments, and 
thank Hidekazu Tanaka, Taku Takeuchi and Takayuki Muto 
for valuable suggestions.
This work was supported by Grant-in-Aid for JSPS Fellows (2008528).
 
{}
  
\end{document}